\documentclass{article}

\usepackage{arxiv}

\usepackage[utf8]{inputenc} 
\usepackage[T1]{fontenc}    
\usepackage{hyperref}       
\usepackage{url}            
\usepackage{booktabs}       
\usepackage{amsfonts}       
\usepackage{nicefrac}       
\usepackage{microtype}      
\usepackage{lipsum}
\usepackage{amssymb}
\usepackage{amsmath}
\usepackage{graphicx}
\usepackage{caption}
\usepackage{multirow}
\usepackage{booktabs}
\usepackage{tabularx}

\title{\textbf{Human Mobility Disproportionately Extends PM\textsubscript{2.5} Emission Exposure for Low Income Populations}}

\author{
    \sffamily\large \vspace{0.15in}
    Chao Fan\textsuperscript{1,*}, Yu-Heng Chien\textsuperscript{2}, Ali Mostafavi \textsuperscript{1,*}\\
    \sffamily\normalsize
    \textsuperscript{1}Department of Civil and Environmental Engineering, Texas A\&M University, College Station, TX, 77843, U.S.\\
    \sffamily\normalsize
    \textsuperscript{2}Department of Industrial and Systems Engineering, Texas A\&M University, College Station, TX, 77843, U.S.\\
    \sffamily\normalsize
    \textsuperscript{*}Corresponding authors: chfan@tamu.edu, amostafavi@civil.tamu.edu\\
}

\begin{document}
\maketitle

\begin{abstract}
\sffamily Ambient exposure to fine particulate matters of diameters smaller than 2.5$\mu m$ (PM\textsubscript{2.5}) has been identified as one critical cause for respiratory disease. Disparities in exposure to PM\textsubscript{2.5} among income groups at individual residences are known to exist and are easy to calculate. Existing approaches for exposure assessment, however, do not capture the exposure implied by the dynamic mobility of city dwellers that accounts for a large proportion of the exposure outside homes. To overcome the challenge of gauging the exposure to PM\textsubscript{2.5} for city dwellers, we analyzed billions of anonymized and privacy-enhanced location-based data generated by mobile phone users in Harris County, Texas, to characterize the mobility patterns of the populations and associated exposure. We introduce the metric for exposure extent based on the time people spent at places with the air pollutant and examine the disparities in mobility-based exposure across income groups. Our results show that PM\textsubscript{2.5} emissions disproportionately expose low-income populations due to their mobility activities. People with higher-than-average income are exposed to lower levels of PM\textsubscript{2.5} emissions. These disparities in mobility-based exposure are the result of frequent visits of low-income people to the industrial sectors of urban areas with high PM\textsubscript{2.5} emissions, and the larger mobility scale of these people for life needs. The results inform about environmental justice and public health strategies, not only to reduce the overall PM\textsubscript{2.5} exposure but also to mitigate the disproportional impacts on low-income populations. The findings also suggest that an integration of extensive fine-scale population mobility and pollution emissions data can unveil new insights into inequality in air pollution exposures at the urban scale. 
\end{abstract}

\keywords{\sffamily Disparity \and Air pollution \and Hazard Exposure \and Mobility \and Environmental Justice \and Equity}

\vspace{0.3in}
\newpage
\section*{\sffamily Introduction}
\vspace{-0.5em}
Fine particles with a mass median aerodynamic diameter of less than 2.5$\mu m$ (PM\textsubscript{2.5}) that can penetrate deeply into the lung and damage the human respiratory system have been identified as a major public health concern \cite{xing8,Kioumourtzoglou2016,T2014}. The World Health Organization (2016) attributes an estimated 4.2 million deaths to exposure to ambient PM\textsubscript{2.5} in air pollution \cite{Organization2016}. Low-income populations are at higher risk of death from being exposed to PM\textsubscript{2.5} \cite{Jbaily2022}. Hence, understanding the exposure of different population groups to PM\textsubscript{2.5} air pollutants is imperative to inform policy interventions not only to reduce the overall exposure but also to provide all people with the same degree of protection from environmental risks \cite{Caplin2019}.

Recognizing the significance of PM\textsubscript{2.5} emission exposure, researchers have devoted significant efforts to measuring disparities in the exposure of PM\textsubscript{2.5} emissions among social-demographic groups at the local, regional \cite{E2021}, and national levels \cite{W2022,Browning210}. Traditional understanding of disparate exposure, however, is largely based on empirical models and air quality data detected by environmental sensors that estimate the concentrations of air pollutants in different areas of a city \cite{Bell2012}. The exposure to air pollutants is often approximated by the concentrations sampled at individual residences and is sensitive to the resolution of demographic data. For example, a prior study assigned the emissions of PM\textsubscript{2.5} from industrial facilities to nearby census block groups and then incorporated this pollutant into calculations of the burden of these emissions on racial groups and by poverty status \cite{Mikati2018} in the adjacent census block. As emissions of PM\textsubscript{2.5} are highly associated with land use and locations of industrial facilities, estimating the exposure based on the emissions from nearby facilities is a simplified and proper way to examine the disparities. To obtain a finer-grained resolution of PM\textsubscript{2.5} concentrations, a number of empirical models have been developed to calculate the spatial gradients of the concentrations. The InMAP air quality model is a commonly adopted model used to estimate concentrations of air pollutants resulting from anthropogenic emissions \cite{Tessum2017}. Then emissions from the source facilities are allocated to individual grid cells within a city using spatial surrogates \cite{W2019}. More recently, advanced technologies such as street-view vehicles equipped with mobile platforms have been employed to measure outdoor air pollutants in neighborhoods \cite{Apte2017}. Researchers could describe such outdoor concentration as “exposure” for block residents \cite{E2021}. These methods from prior studies allow us to estimate fine-scale resolutions of PM\textsubscript{2.5} concentrations at individual residences. However, exposure to PM\textsubscript{2.5} emissions is also influenced by dynamic human activities—work and life activities—that occur mostly outside the home, creating variations in the exposure to ambient emissions \cite{Fan2021,fan2021neural}. 

The significance of human dynamics in environmental justice problems has become a matter of concern. An increased number of recent studies have attempted to process human-generated data and capture dynamic mobility patterns of populations \cite{Deng2021,Fan2022}. In particular, the ubiquity of smartphones and their embedded technology for location collection offers unique opportunities for research and practices \cite{Maas2019}. Mobile phone data that captures the timestamped locations of anonymized devices capture the trajectories of individuals, population density at places \cite{Deville2014}, and visit patterns of populations \cite{Blondel2015}. One of the most popular metrics that prior research has developed and been commonly adopted to characterize the patterns of human mobility is the radius of gyration \cite{Song2010}, interpreted as the characteristic distance traveled by an individual in a given time window. This metric has been successfully integrated into statistical models to predict individual mobility \cite{Gonzalez2008} and crowd movements \cite{Yan2017a}. The results from existing literature \cite{Widhalm2015} show the suitability of this metric in describing the mobility scales of individuals. By integrating radius of gyration with social and physical contexts in a city, some studies \cite{Pan2013} investigated visit patterns extracted from population mobility, such as individual lifestyles based on the sequences of visited places \cite{DiClemente2018}. These studies overlay mobile phone data onto the points of interest data that document attributes of places, such as names, brands, and sectors \cite{Zeng2017,yao2018}. Specifically, this contextual information enables the capture of frequently visited places of individuals and the length of stay at the places to uncover the daily life needs and work activities of city dwellers \cite{Ren2014}. In short, the gyration radius and contextual information in analytics of mobile phone data have promoted a variety of applications, including the examination of disparities in disaster evacuation by race and income \cite{Deng2021}, quantification of income segregation during social interactions \cite{Moro2021}, and characterization of urban structures with social and demographic inequalities \cite{fan2021neural}. These studies \cite{Gao2020,Alexander2015,barbosa2018human}, however, have not considered the intersection between human dynamics and the environment to understand how mobility influences the exposure of the populations to PM\textsubscript{2.5} air pollutants. 

In this study, using a large collection of fine-scale mobility data generated by mobile phones, we address how mobility behaviors extend the exposure of populations to PM\textsubscript{2.5} emissions. This study integrates human dynamics into exposure measurement. We introduce a time-weighted and location-associated exposure metric to examine the time people spend at urban places and the emissions released by nearby industrial facilities. As a testbed, we selected Harris County in the Houston metropolitan area, the third most populous US county, whose population has a wide range of household incomes \cite{Bureau}. As the Energy Capital of the World, Harris County is a diverse and vibrant metro area that is home to a concentration of the oil and gas industries, chemical and allied product manufacture, and petroleum and related industrial processes. A consequence of prosperity in the oil and gas industry is a steep increase in the emissions of air pollutants, which may increase health risks for city dwellers from routine exposure to these pollutants. Hence, Harris County could serve as a suitable and representative study area to measure and compare the exposure of urban dwellers to air pollutants. The implementation of the metrics in Harris County reveals the disparities in mobility-based exposure across urban income groups. Furthermore, we identify residential and mobility features associated with the disparate exposure, which provide important insights related to the drivers of the disparity and could inform targeted policymaking for not only reducing the overall PM\textsubscript{2.5} exposure but also providing all populations with the same level of protection from the PM\textsubscript{2.5} risks in an equitable manner.

\section*{\sffamily Methods}
\vspace{-0.5em}
\subsection*{\sffamily Estimate PM\textsubscript{2.5} Concentration}
\vspace{-0.7em}
We used the 2017 air emissions inventory data released by the United States Environmental Protection Agency (USEPA) in 2021 to estimate the presence of air pollutants in Harris County. The National Emissions Inventory (NEI) data \cite{emissions2017} includes a facility-level inventory that summarizes the emissions of a variety of air pollutants, and locations and sectors of specific facilities which discharge air pollutants in industrial processes. This study specifically focused on air pollutants for PM\textsubscript{2.5}, one of the main causes of respiratory diseases. To estimate the concentration levels of air pollutants for PM\textsubscript{2.5} across the urban areas, we established a grid map dividing the Houston metro area of the city into 4280 1-km x 1-km (approximate) grid cells. Air pollutants emitted from the facilities within a grid cell account for the total concentration of PM\textsubscript{2.5} in the grid cell. Based on detailed geographical information of the facilities emitting air pollutants for PM\textsubscript{2.5}, we calculated the PM\textsubscript{2.5} concentration by summing up the emissions from all facilities within the grid cell.

\subsection*{\sffamily Measure Experienced Exposure}
\vspace{-0.7em}
The experienced exposure to PM\textsubscript{2.5} is estimated from the mobility activities of the populations. Here, we capture the mobility activities from anonymized and aggregated location data of mobile phone devices. The data set is provided by Spectus Inc., a location data intelligence company which collects anonymous, privacy-compliant location data of mobile devices using their software development kit (SDK) technology in mobile applications and ironclad privacy framework \cite{Aleta2020a}. The Spectus techniques process data only from mobile devices whose owners have opted in to share their location and require all application partners to disclose their relationship with Spectus, directly or by category, in the privacy policy. With this commitment to privacy, the data set contains location data for roughly 15MM daily active users in the United States. Through Spectus’ Data for Good program \cite{Spectus}, Spectus provides mobility insights for academic research and humanitarian initiatives. All data analyzed in this study are aggregated to preserve privacy. Each entry in the data table comprises anonymized device ID, coordinates, start time, and dwell time of the stop for the device, as well as the points of interest associated with the stop. In order to further preserve privacy, Spectus also obfuscates home areas by estimating the census block group (CBG) where the mobile devices stayed during nighttime hours. The representativeness of this data has been demonstrated by prior studies \cite{Aleta2020a,Nande2021}.

\begin{figure}
  \centering
  \includegraphics[width=17cm]{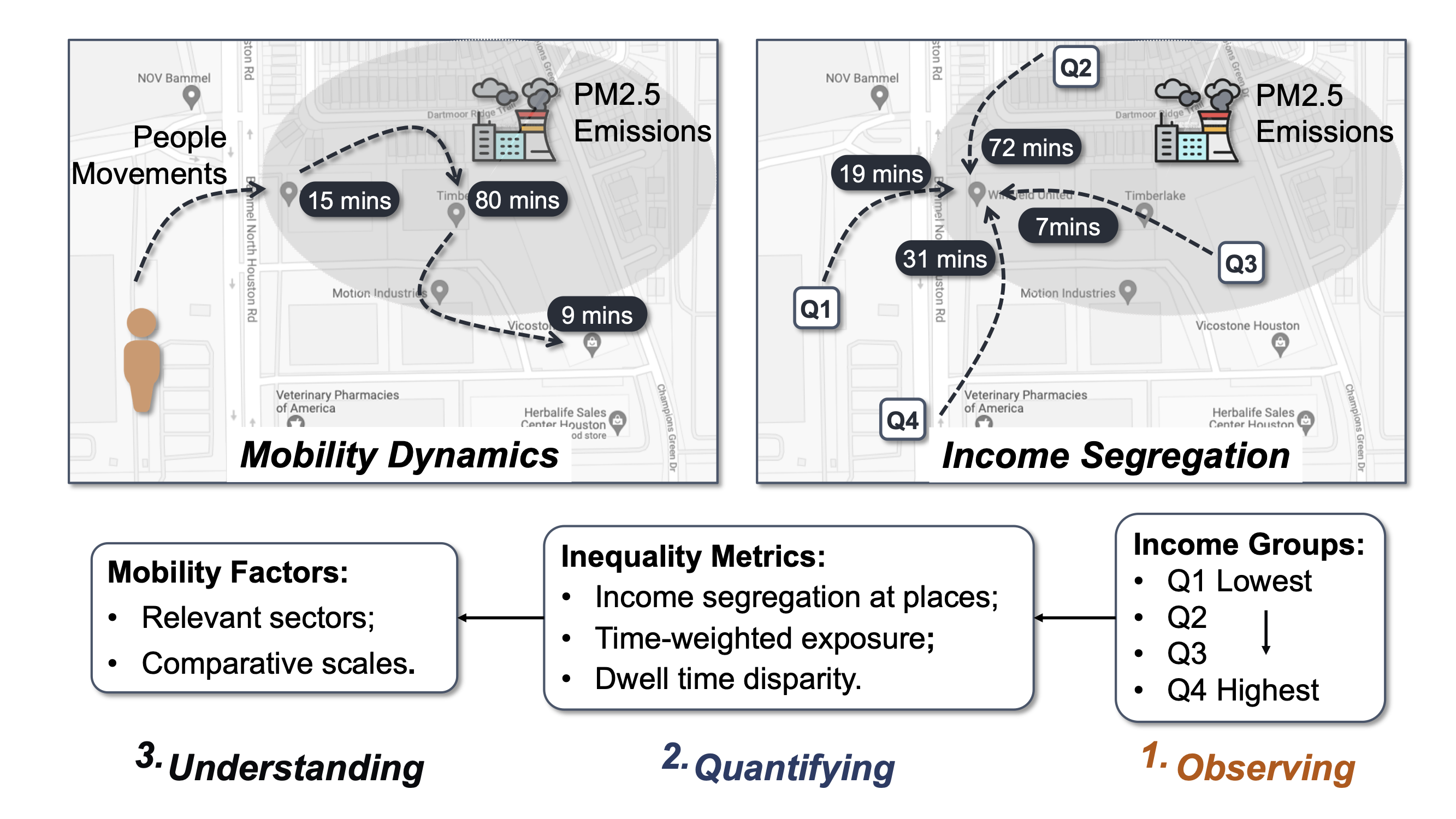}
  \caption{\textbf{Mobility dynamics and income segregation.} (Synthetic user trace displayed in visualization to preserve privacy for demonstrative purposes.) The schematic illustration shows the development of the metrics and implementation of analyses in this study. We associated the median household income of a neighborhood with the residents living in the corresponding neighborhood. Based on their associated income, residents are then classified into four groups of equal size. The module on mobility dynamics proves the fact that people move along a trajectory and spend time visiting places of interest. The dwell time that a person spends in a place of interest varies across the places on the trajectory. With the associated income groups of visitors, the module of income segregation indicates that people who visited the same place may be from different income groups. This concept motivates the development of our metrics for emission exposure. Finally, the study reveals relevant factors for understanding the mechanisms of mobility patterns and emission exposure. }
  \label{fig:fig1}
\end{figure}

We upscaled each stop of the mobile devices to a grid cell in Harris County in Texas. The trajectories of the mobile devices are then converted into a set of grid cells capturing the dwell time in each grid cell. The plot in the top right corner of \textbf{Fig. \ref{fig:fig1}} shows the mobility dynamics of the population captured by our method and data set. The spatial resolution of the mobility activities of the population becomes consistent with the spatial resolution we estimated for PM\textsubscript{2.5} concentration. Such spatial consistency allowed us to superimpose one data layer over the other to measure mobility-based exposure of populations to PM\textsubscript{2.5}. We define a time-weighted PM\textsubscript{2.5} concentration as the exposure of a mobile device. Specifically, the experienced exposure to PM\textsubscript{2.5} is associated with the PM\textsubscript{2.5} concentration in a grid cell and the device’s dwell time in the cell. Hence, the time-weighted PM\textsubscript{2.5} concentration e, which an individual $i$ is exposed to, is given by:

\begin{equation}
    e_i=\sum_{\alpha}^{n}{\tau_{i\alpha}\cdot\psi_\alpha}
\end{equation}

where $\tau_{i\alpha}$ denotes the time the individual $i$ spent at the grid cell $\alpha$, $\psi_\alpha$ denotes the PM\textsubscript{2.5} concentration at the grid cell $\alpha$, and $n$ is the total number of grid cells visited by that individual. As such, the measurement of experienced exposure to PM\textsubscript{2.5} concentration for a neighborhood (same as CBG in this study) can be obtained by averaging the exposure of all individuals whose homes are in the same neighborhood. Here, we define, $\kappa_\beta$, the average time-weighted exposure of people from a CBG $\beta$ as:

\begin{equation}
    \kappa_\beta=\frac{1}{N_\beta}\sum_{i}^{N_\beta}\sum_{\alpha}^{n}{\tau_{i\alpha}^{\left(\beta\right)}\cdot\psi_\alpha}
\end{equation}

where $N_\beta$ denotes the number of individual devices from CBG $\beta$. This measurement serves as the strategy to upscale the experienced exposure to PM\textsubscript{2.5} from individuals to neighborhoods, which reveals the extent of experienced exposure to air pollutants for a group of people with a similar social and economic status.

\subsection*{\sffamily Measure Income Segregations}
\vspace{-0.7em}
The disparities in air pollutant exposure are the result of segregated visits from different population groups to emission places. A population group can be denoted, for instance, as those in the first decile of an income group. The variable of interest in this study is the median household income of a population group. We obtained this economic variable for CBGs in Harris County, Texas, from the 2014–2018 five-year American Community Survey (ACS) conducted by the US Census Bureau \cite{USCensusBureau2019}. We assigned all individual devices with the median household income based on their estimated home CBGs and categorized them into four income groups of equal population size. Throughout the paper, we use Q1 to Q4 to label the income groups from the lowest to the highest, respectively. 

The plot on the top left corner of \textbf{Fig. \ref{fig:fig1}} illustrates the segregated visit patterns of people to different places. We quantified income segregation at places based on data from an existing study \cite{Moro2021}, first defining a metric, $\theta_{q\alpha}$, which indicates the proportion of average time an individual from income quantile $q$ spent at a grid cell $\alpha$. The metric is formulated as follows:

\begin{equation}
    \theta_{q\alpha}=\frac{\frac{1}{N_{q\alpha}}\sum_{i}^{N_{q\alpha}}\tau_{iq}^{\left(\alpha\right)}}{\sum_{q=1}^{4}\left[\frac{1}{N_{q\alpha}}\sum_{i}^{N_{q\alpha}}\tau_{iq}^{\left(\alpha\right)}\right]}
\end{equation}

where $\tau_{iq}^{\left(\alpha\right)}$ denotes the time an individual $i$ from income quantile $q$ spent at a grid cell $\alpha$, and $N_{q\alpha}$ denotes the total number of individuals from income quantile $q$ visiting the grid cell $\alpha$. This metric quantifies the visit and dwelling patterns of people from different income groups at specific places (grid cells). Then the income segregation within a grid cell is measured by the deviation from a balanced mixing of income groups based on the average amount of time they spent. The measure, $S_\alpha$, is defined as:

\begin{equation}
    S_\alpha=\frac{2}{3}\sum_{q}\left|\theta_{q\alpha}-\frac{1}{4}\right|
\end{equation}

when people from different income quantiles visited the grid cell $\alpha$ with equality of time, $\theta_{q\alpha}=1/4$ for all income quantiles and $S_\alpha=0$. When people’s visit time is highly unequal across different income quantiles at the grid cell $\alpha$, $\theta_{q\alpha}$ could be extremely close to 0 or 1, and $S_\alpha$ would be close to 1. Hence, the measure $S_\alpha$ is bounded between 0 and 1, and a smaller value of $S_\alpha$ represents a more equal visit pattern at the grid cell $\alpha$. This metric enables the quantification of income segregation at places and provides a rationale for the unequal air pollutant exposure of people from different income quantiles at places.

\subsection*{\sffamily Characterize Mobility Activities}
\vspace{-0.7em}
The disparities in experienced exposure to air pollutants among income population groups are the outcome of the interactions between mobility activities of humans and the places of the emission sources. Understanding such interactions is critical for the mitigation of the extent and disparities of experienced exposure. Two dimensions in the mobility activities are associated with the emission places: the sectors of the emission places where people spent the most of their time; and the scale of mobility, indicating the extent to which the scale of people’s activities have overlaps with places of emission sources. Hence, we propose appropriate metrics to characterize the mobility activities in both dimensions. 

The first metric is defined to measure the extent to which the time spent at a sector of facilities by people differs across different income quantiles. Here, a sector of facilities means the type of facilities, which is regulated by the Multi-Sector General Permit. The US Environmental Protection Agency delineates 12 sectors for emission facilities among 29 industrial sectors. We used the deflection of time, $\delta_{sector}$, spent at an emission sector by individuals from income quantile $q$ to characterize the patterns of mobility activities. The metric is computed as follows:

\begin{equation}
    \delta_{sector}=\frac{1}{N}\sum_{\gamma\in\left\{sector\right\}}\delta_{q\gamma}
\end{equation}

where $N$ represents the total number of people visiting the emission sectors. $\delta_{q\gamma}$  measured the deflection of time spent at facility $\gamma$ by individuals from income quantile $q$:

\begin{equation}
    \delta_{q\gamma}=\frac{\frac{1}{N_{q\gamma}}\sum_{i}^{N_{q\gamma}}\tau_{iq}^\gamma-e}{e}
\end{equation}

\begin{equation}
    e=\frac{1}{N_\gamma}\frac{1}{N}\sum_{q=1}^{4}\sum_{\gamma}^{N_\gamma}\sum_{i}^{N_{q\gamma}}\tau_{iq}^\gamma
\end{equation}

where $\tau_{iq}^\gamma$ represents the time an individual $i$ from income quantile $q$ spent at the facilities in a sector $\gamma$, $N_{q\gamma}$ represents the total number of people from income quantile $q$ visited facilities in sector $\gamma$, and $N_\gamma$ represents the total number of documented emission sectors. 

The second metric we propose to characterize mobility activities is the mobility scales of people from their home grid cells to all other visited grid cells. Specifically, we adopted the idea of the radius of gyration for the mobility activities of each individual. The mobility radius of an individual $i$ from a neighborhood $\beta$ is calculated based on the distances $d_{i\alpha}$ from the home grid cell to the visited grid cells $\alpha$. The equation is shown as follows:

\begin{equation}
    m_i^2=\frac{1}{N_\alpha}\sum_{\alpha}^{N_\alpha}d_{i\alpha}^2
\end{equation}

where $N_\alpha$ is the total number of grid cells an individual $i$ has visited. Then, the scale of mobility for the people from a specific neighborhood can be calculated as:

\begin{equation}
    m_\beta=\frac{1}{N_\beta}\sum_{i}^{N_\beta}m_i
\end{equation}

where $N_\beta$ is the total number of individuals living in the neighborhood $\beta$. 

To examine the overlap between the scales of mobility activities and the places of emission facilities, we also measured the average radial distance from a neighborhood $\beta$ to emission facilities. 

\begin{equation}
    M_\beta^2=\frac{1}{N_\alpha}\sum_{\alpha}^{N_\alpha}d_{\beta\alpha}^2
\end{equation}

where, $d_{i\alpha}$ represents the distance from neighborhood $\beta$ to the grid cell $\alpha$. This metric will shed light on how close the neighborhood is located to emission facilities.

\section*{\sffamily Results}
\vspace{-0.5em}
\subsection*{\sffamily Patterns of PM\textsubscript{2.5} Concentrations and Segregation}
\begin{figure}
  \centering
  \includegraphics[width=17cm]{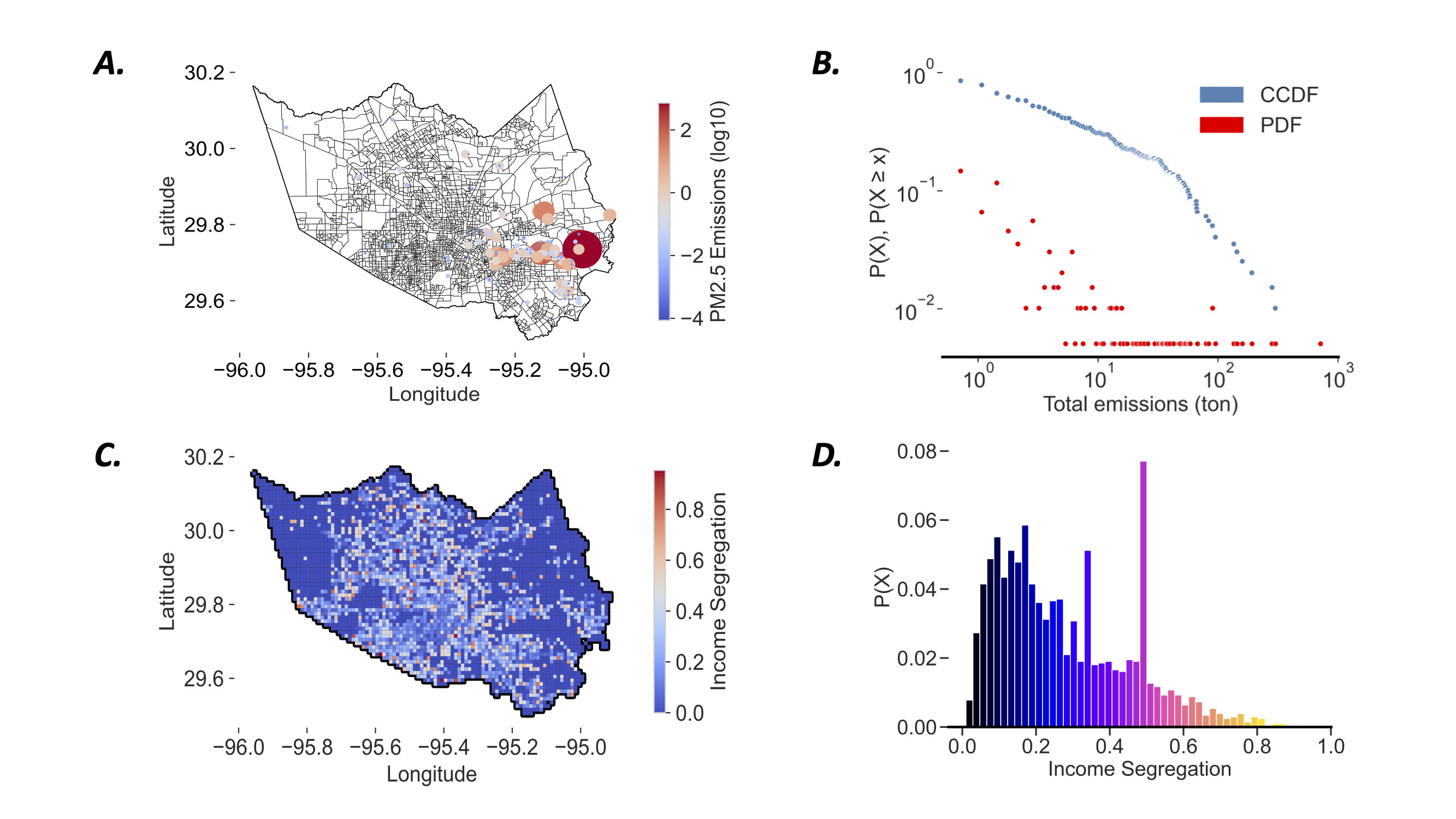}
  \caption{\textbf{Distribution of air pollutant (PM\textsubscript{2.5}) emissions and experienced income segregation.} {\sffamily \textbf{A.}} Geographical distribution of air pollutant emissions by grid cell across Harris County. The magnitude of emissions is in the logarithmic scale and is shown in a color bar. The size of the ball represents the magnitude of the emissions. {\sffamily \textbf{B.}} Distribution of the total emission intensity (ton). PDF is probability density function; and CCDF is complementary cumulative density function. Both axes are in log10 scale. {\sffamily \textbf{C.}} Mapping of the income segregation in grid cells in Harris County, Texas. The values of the metric range from 0 to 1, distinguished by color. The higher value of income segregation in a grid cell indicates a highly unequal amount of time that people from different income groups spent in the grid cell. {\sffamily \textbf{D.}} Probability mass function (PMF) for income segregation in grid cells in Harris County showing the distribution of the extent of segregation across the county. Note that grid cells with no visits are excluded from the distribution plot.}
  \label{fig:fig2}
\end{figure}

\vspace{-0.7em}
Air pollutant-emitting facilities are unevenly distributed throughout the urban areas of Harris County; their emissions vary significantly, leading to uneven concentrations of PM\textsubscript{2.5} in different areas of the county. \textbf{Fig. \ref{fig:fig2}A} maps the concentrations of PM\textsubscript{2.5} emissions in Harris County by aggregating the emissions from all facilities in each grid cell. The county has many areas with low concentrations of PM\textsubscript{2.5}, while high-intensity PM\textsubscript{2.5} is concentrated in a few areas, such as the eastern part of the county. Such a spatial pattern of pollutant concentrations indicates a heavy-tailed probability distribution function, as shown in \textbf{Fig. \ref{fig:fig2}B}. Of the grid cells in Harris County, 1\% of the grid cells emit more than 100 tons/year PM\textsubscript{2.5}. The concentrations of PM\textsubscript{2.5} in the rest of the grid cells range from 0 to 10 tons/year. This highly unequal distribution of quantities and locations of PM\textsubscript{2.5} concentrations suggests the potential of unequal exposure of populations to PM\textsubscript{2.5} pollution based on their activities in different areas of the county. 

Due to dynamic human activities, exposure to emissions extends beyond the location of residence. People from different income groups may exhibit diverse activity patterns in different urban areas. We employed a segregation metric for each grid cell to measure the degree of segregation of the visited places in terms of the activity patterns of different income groups. The segregation metric captures the time spent in each grid cell by different income groups and the extent to which the time spent varies across different income groups. \textbf{Fig. \ref{fig:fig2}C} shows the results of measuring the experienced income segregation of grid cells in Harris County. The majority of urban areas in Harris County are income segregated. That is, the amount of time residents from different income groups spent in these grid cells is significantly unequal. Income segregation is pronounced, especially in grid cells at the center of the county. The extremely segregated grid cells are dispersed across the county and surrounded by moderately and mildly segregated grid cells. Those extremely segregated grid cells may only be visited by people from a single income group, such as high-expense places expenses or places where only a particular income group population would likely work. 

The distribution for the time-spent income segregations in grid cells is plotted in \textbf{Fig. \ref{fig:fig2}D}. Most of the segregated grid cells tend to have 0 to 0.5 deflection from equality of visit time among income groups. Extremely segregated areas, such as the grid cells with income segregation greater than 0.6, are observed and account for 1\% to 2\% among the grid cells in the county. By juxtaposing the time-spent income segregation and PM\textsubscript{2.5} emission maps, we find that areas with high PM\textsubscript{2.5} concentrations are overlapping with the areas with significant time-spent income segregation. This result confirms that the interactions between mobility activities and emission distributions 

\subsection*{\sffamily Disparities in Mobility-based Exposure}
\begin{figure}
  \centering
  \includegraphics[width=17cm]{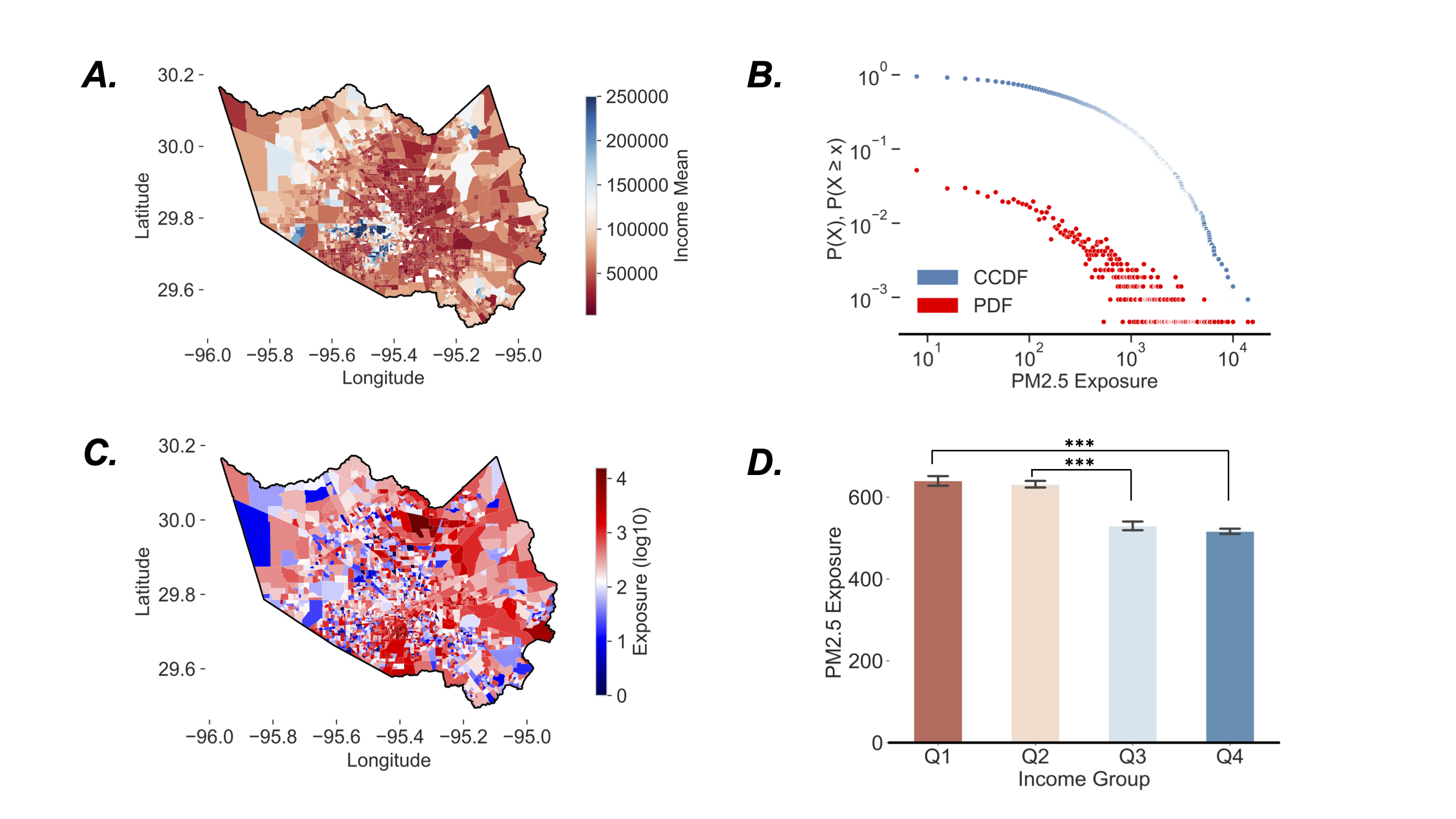}
  \caption{\textbf{Disparities of mobility-based exposure of PM\textsubscript{2.5} due to mobility.} {\sffamily \textbf{A.}} Mean household income of census block groups (CBG) in Harris County, Texas. The polygon and income data are obtained from US Census survey data. {\sffamily \textbf{B.}} Probability density function (PDF) and complementary cumulative density function (CCDF) of the experienced exposure to PM\textsubscript{2.5}. The value of exposure is calculated by the sum of concentrations (ton) at visited grid cells. Both x and y axes are in the log10 scale. {\sffamily \textbf{C.}} Mobility-based exposure to PM\textsubscript{2.5} at CBG level. This CBG-level exposure is the average of the individual mobility-based exposure from a CBG, using equation (2). Since the exposure varies significantly, we transformed the values into a log10 scale on the map. {\sffamily \textbf{D.}} Exposure distribution of four income groups in Harris County. The error bar shows 95\% confidence intervals around the mean value on each bar. Q1 is the lowest-income group; Q4, the highest-income group. The colors representing income groups are consistent with the colors in the subplot A. Significance tests are also conducted to examine the significance of the differences in mobility-based exposure among income groups. Note: $*** p \leq 0.01$; $** p \leq 0.05$; $* p \leq 0.1$. The detailed results for the significance tests are provided in the Supplementary Information Table 1.}
  \label{fig:fig3}
\end{figure}

\vspace{-0.7em}
Integrating the mobility activities of people at places into the exposure metric, the results reveal the patterns of disparate exposure to PM\textsubscript{2.5} among urban populations. \textbf{Fig. \ref{fig:fig3}B}. shows a heavy-tailed distribution of the mobility-based exposure to PM\textsubscript{2.5} for the neighborhoods (CBGs) in Harris County. This result indicates that, although a large number of neighborhoods are exposed to pollutants with a light concentration of PM\textsubscript{2.5}, considering mobility activities, a few CBGs suffer severely from the exposure to high concentrations of PM\textsubscript{2.5}. Such distinct disparities of exposure unveil an important environmental injustice issue related to how different income groups are placed in this distribution. 

We mapped both the mean household income of CBGs (\textbf{Fig. \ref{fig:fig3}A}.) and exposure of CBGs to PM\textsubscript{2.5} (\textbf{Fig. \ref{fig:fig3}C}.) to examine how the income groups and the air pollutant exposure are spatially related. On one hand, according to \textbf{Fig. \ref{fig:fig3}A}., we find that the high-income populations are concentrated in the southwest of the county, while low-income populations are located from the center to the east and north of the county. On the other hand, the map of mobility-based exposure at the CBG level in \textbf{Fig. \ref{fig:fig3}C}. shows that PM\textsubscript{2.5} exposure is pervasive throughout the urban populations. All neighborhoods are more or less exposed to emissions of PM\textsubscript{2.5} due to population mobility activities. In particular, residents of the west and south of the county are less exposed to PM\textsubscript{2.5} based on their mobility activities, compared to residents of the north and east of the county. This qualitative spatial understanding of the interactions between income groups and mobility-based air pollutant exposure implies their association underneath the overall exposure disparities. 

To future examine the significance of this finding, based on the classification of income groups we defined previously, \textbf{Fig. \ref{fig:fig3}D}. shows how different the mobility-based exposure is across urban income groups. Low-income population groups (Q1 and Q2 groups) have a significantly higher exposure to PM\textsubscript{2.5} based on their mobility activities, compared with the exposure of high-income population groups (Q3 and Q4). This result indicates a disproportionate exposure of low-income populations to air pollutants (PM\textsubscript{2.5}) in the city. This inequality of mobility-based exposure to the emissions is greater than exposure determined based on traditional approaches.

\subsection*{\sffamily Disparities Relative to Mobility Patterns}
\vspace{-0.7em}
The spatial and quantitative understanding of disparate exposure among income groups reveals an important phenomenon in which the concentrations of PM\textsubscript{2.5} disproportionally impact low-income populations due to their mobility activities as well as to the location of pollutant-emitting facilities. This mobility-based measure of exposure relies on specific places people visit, the concentration of PM\textsubscript{2.5} at these places, and the time people spent at these places. To deepen the understanding of the mechanisms of disparate mobility-based exposure, we investigate the patterns of mobility activities of different income population groups and how these patterns are associated with the emissions of air pollutants. In this section, we mainly examine two relationships: (1) the relationship between the sectors of the emissions and the mobility activities of people in areas where these sectors are located to understand the impacts of visit patterns on air pollutant exposure; and (2) the relationship between scales of mobility and the distances from home neighborhoods to emission locations, to capture the spatial reach and overlaps of population activities and the emission sources. 

\begin{figure}
  \centering
  \includegraphics[width=17cm]{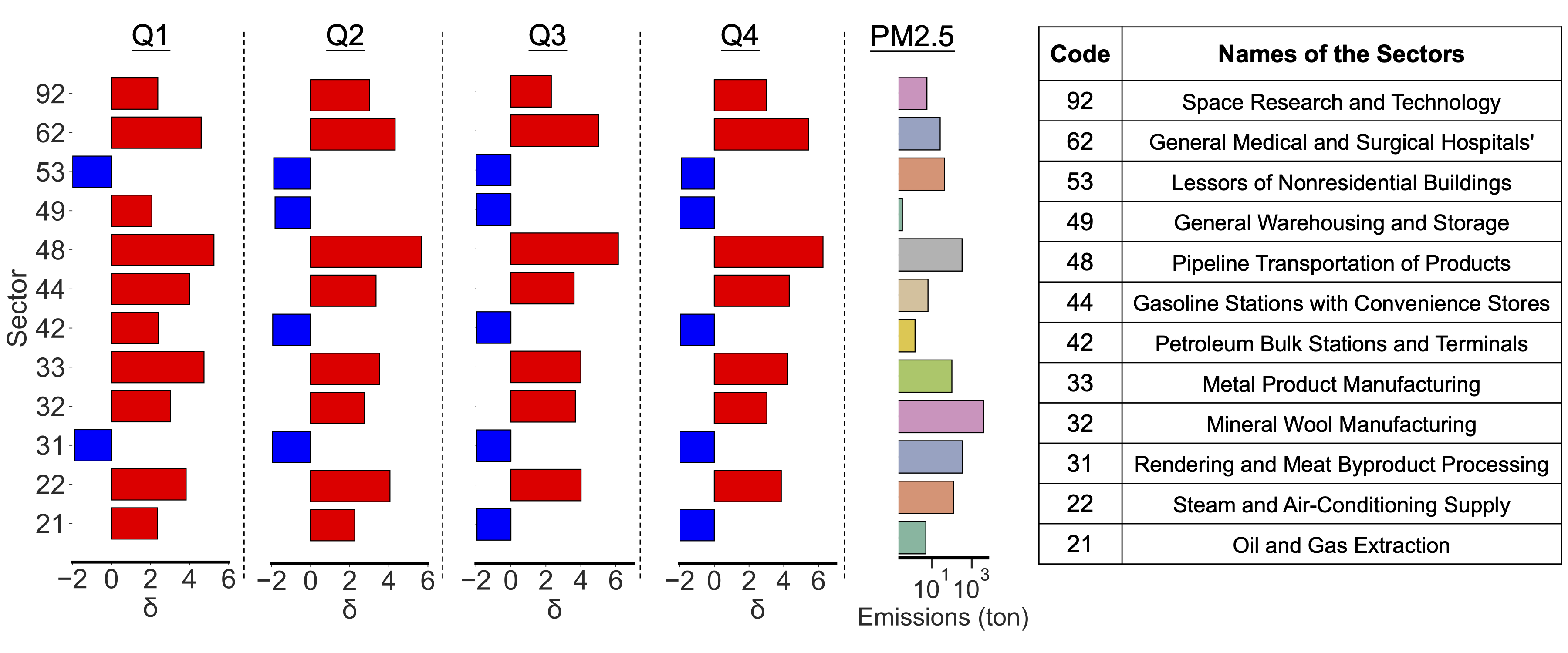}
  \caption{\textbf{Relationship between emission sectors and mobility patterns of the populations.} The PM\textsubscript{2.5} concentrations and the time spent by people at places are aggregated by the sectors of the emission sources. $\delta$(\%), here, denotes the deflection of time spent at facilities in a specific sector by individuals from an income group. The ticks and numbers on the axis for the first four subplots are in the log10 scale. According to the average time spent by the individuals at all places in a county, the metric $\delta$ could show the negative and positive deflection of the time spent by the population groups in the sectors. The first four subplots from the left show the deflection of time for four income groups.  Q1 to Q4 represent four income groups based on their mean household incomes from low to high. The rightmost subplot describes the PM\textsubscript{2.5} emissions of each sector based on EPA data. A table of the sectors’ codes and descriptions sectors on the right of the plot. A complete table showing the names of the sectors is provided in the Supplementary Information.}
  \label{fig:fig4}
\end{figure}

The results in \textbf{Fig. \ref{fig:fig4}} reveal that the times people from different income groups spent in the concentration areas are similar, such as the time spent in the space research and technology sector and medical and hospitals sector. This phenomenon is dominant in most sectors; however, in a few sectors, the difference in the time people spent varies among different income groups, which shows an underlying mechanism for disparate mobility-based exposure to PM\textsubscript{2.5}. Specifically, the greater exposure of low-income populations is a result of the greater-than-average time they spent in these emission sectors, including the oil and gas extraction sector, petroleum bulk stations and terminals sector, and warehousing and storage sector. This observation indicates that low-income people may have a higher probability to work or interact with facilities such as mining, oil and gas extraction, compared with high-income populations. Such mobility patterns of the low-income people result in greater exposure of these sub-populations to PM\textsubscript{2.5} in urban areas.

\begin{figure}
  \centering
  \includegraphics[width=17cm]{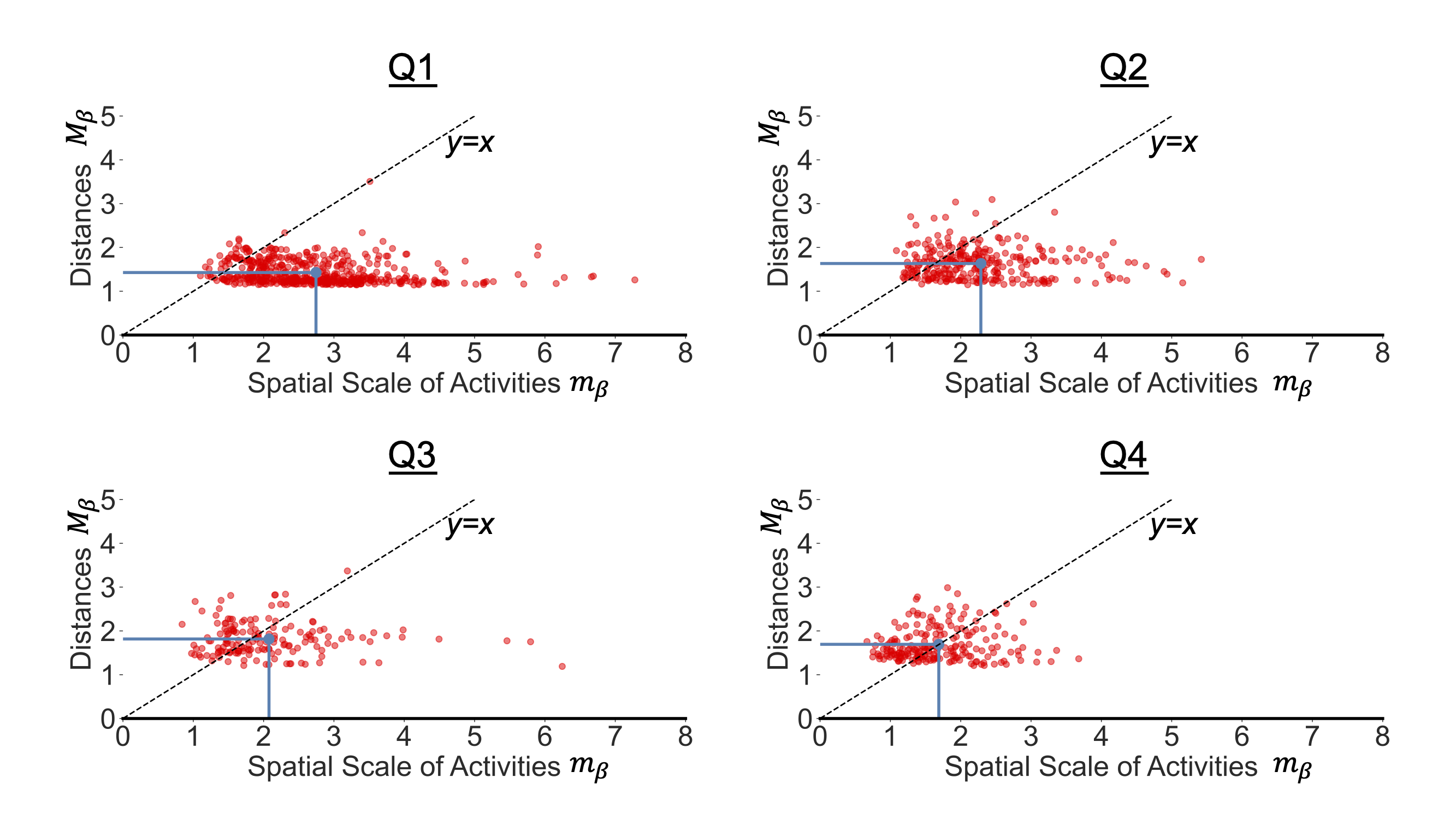}
  \caption{\textbf{Relationship between the scales of mobility activities and the distances to emission facilities.} Both metrics are estimated at the neighborhood (CBG) level. Each subplot represents the relationships for an income population group. Q1 through Q4 represent four income groups based on their mean household incomes from low to high. The red dots in the plots represent the neighborhoods in Harris County. The blue vertical and horizontal lines are the mean values of the distances $M_\beta$ and mobility scale $m_\beta$. The intersection of these two lines could show the difference between the distances to emission facilities and the scale of population mobility in the income group. The dark dot line of $y=x$ supports the evaluation of differences between the two metrics. The numbers on both the $x$ and $y$ axes in the subplots are in kilometers.}
  \label{fig:fig5}
\end{figure}

In addition to the exposure of people in the work environment, the association between the visit patterns of people for their life activities and the exposure to PM\textsubscript{2.5} is another important driver of exposure. The scale of an individual’s mobility has been studied for decades and is considered a key measurement of mobility patterns \cite{Gonzalez2008}. In this study, we characterize the scale of mobility for people from the same neighborhood based on the radius of gyration, and we examined the varying relationships between the scales of mobility activities and the distances to emission facilities among income groups. Results depicted in \textbf{Fig. \ref{fig:fig5}}. indicate that the distances from the home neighborhoods to emission facilities are generally consistent across the four income groups; however, the scales of mobility for people from different income groups vary significantly. 

We observed that the average radius of gyration for mobility activities of low-income populations (Q1 and Q2) is even 1 kilometer more than the average radius of gyration of high-income populations (Q3 and Q4). Hence, for low-income populations, the scales of their mobility activities are greater than the distances from their home neighborhoods to emission source facilities, while high-income people tend to have a smaller scale of mobility. A great scale of mobility that comes across a large spatial area would increase the possibilities of exposure to air pollutants. People with higher incomes move shorter distances for life activities, which leaves them less exposed to air pollutants, compared to people with lower incomes. This result raises important implications related to the catchment and accessibility of low-income populations for their life needs. We will discuss these implications in detail in the discussion section.

\section*{\sffamily Discussion and Concluding Remarks}
\vspace{-1em}
It is well known to researchers and practitioners that air pollutants like PM\textsubscript{2.5} have disparate impacts on the respiratory health of urban populations; however, little was known about the extent to which the interweaving of emission sources and mobility behaviors exposes low-income populations disproportionately to PM\textsubscript{2.5} emissions in cities. In this study, we integrate human dynamics into the measurement of air pollution exposure in urban communities. The approach we proposed here leverages a unique and large set of mobile phone data which documents the mobility activities of the populations to capture the time people spent at places surrounding emission facilities. The emission data provided by the US Environmental Protection Agency allows us to estimate the concentration of PM\textsubscript{2.5} at places and calculate the exposure of people to PM\textsubscript{2.5} due to their mobility activities. We quantify the relationships between mobility-based exposure and household incomes. Our results illustrate how detailed and extensive fine-scale mobility data related to human dynamics can add new insights into differences and disparities in air pollution exposures among urban income groups. The methods and findings of this study could inform policymaking to reduce disparities in urban air pollution exposure and the risks to respiratory health. 

First, one core finding is that emissions of PM\textsubscript{2.5} disproportionately impact low-income population groups during their mobility activities. We also observe a huge gap in the levels of exposure between medium-low (Q2) and medium-high (Q3) income groups, while the differences between Q1 and Q2 and the differences between Q3 and Q4 are not so conspicuous. This finding is statistically significant, rendering it improbable that they are attributable to concentration estimation or mobility data biases. This observation highlights that mobility activities of populations have a pronounced impact on the exposure to PM\textsubscript{2.5} between high- and low-income population groups. Hence, in addition to measuring and mitigating the exposure to air pollution based on places of residence \cite{Jbaily2022}, another important disparity, which is in the mobility-based exposure, should be recognized in developing policies for exposure reduction. This study serves as the first attempt to integrate human dynamics into the exposure metric to advance the understanding of exposure disparities among income populations. Based on the findings, we would suggest that more specifically targeted PM\textsubscript{2.5} reduction policies and strategies considering human mobility dynamics are necessary not only to reduce the overall air pollution levels but also to provide all people with a similar degree of protection from environmental and atmospheric hazards. For example, policies can limit urban development around polluting facilities to reduce human activity and mobility surrounding those facilities. The findings could also inform public health policies targeted at reducing respiratory diseases associated with air pollution. 

Second, the problem of disparate exposures to environmental pollution has received growing attention from environmental agencies in the United States. Insights and knowledge that can dive into the drivers of the identified disproportionate impacts of air pollution, however, have been underexplored. Through the investigation of the dynamic human mobility activities, we find the low-income population activities increases exposure to PM\textsubscript{2.5}, including work locations and their mobility scales. In particular, low-income populations tend to work in industries with more PM\textsubscript{2.5} emissions, compared to the PM\textsubscript{2.5} emissions in the working environment of high-income populations. The work environment exposes low-income people more to air pollutants. Also, neighborhoods where the low-income people live may lack sufficient facilities for their life needs, resulting in a long travel distance and a large travel scale in daily mobility. These findings have notable implications for developing effective strategies to reduce both the overall extent and inequality of PM\textsubscript{2.5} exposure among the urban populations. For example, emission reduction policies for the industries where low-income people often work could contribute to reducing the exposure of people during work. In addition, eliminating food deserts in low-income areas by promoting the construction of grocery stores across the city to reduce the mobility scales of the low-income people for life needs could also be helpful for reducing the disproportional impacts of air pollution. 

This study, as a first attempt to integrate human dynamics into environmental justice issues, also comes with some limitations. First, the data available for estimating the PM\textsubscript{2.5} emissions are limited to every three years from 2011 to 2017, released by US EPA. The concentrations and spatial gradients of the air pollutant may differ at depending upon time and location due to changes in atmospheric dynamics. Further research may consider the dynamics of the emissions and atmosphere in the model to improve the measure of disparities present in cities. Second, this study focuses on the exposure disparities of PM\textsubscript{2.5} emissions at a local scale. Pollutant control policies tend to be regional because of the transmission of the air pollutants through the atmosphere. Hence, it is important to consider a larger scale including both urban and rural areas in the analyses to explore the regional disparities in PM\textsubscript{2.5} exposure. Third, although the examination of exposure disparities shown in this study is based on human mobility data and facility emission data from Harris County (Houston), Texas, potential variations might exist in different cities due to the development of the cities and the focuses of the industrial sectors. Experiments can be conducted in other cities to support the generality of our findings from this study.

\section*{\sffamily Acknowledgement}
\vspace{-0.5em}
This material is based in part upon work supported by the National Science Foundation under Grant CMMI-1846069 (CAREER), the Texas A\&M University X-Grant 699, and the Microsoft Azure AI for Public Health Grant. The authors also would like to acknowledge the data support from Spectus. Any opinions, findings, conclusions, or recommendations expressed in this material are those of the authors and do not necessarily reflect the views of the National Science Foundation, Texas A\&M University, Microsoft Azure or Spectus Inc.

\section*{\sffamily Competing interests}
\vspace{-0.5em}
The authors declare that there are no competing interests.

\section*{\sffamily Data availability}
\vspace{-0.5em}
All data were collected through a CCPA- and GDPR-compliant framework and utilized for research purposes. The data that support the findings of this study are available from Spectus Inc., but restrictions apply to the availability of these data, which were used under license for the current study. The data can be accessed upon request submitted on Spectus.ai. Other data we use in this study are all publicly available. 

\section*{\sffamily Code availability}
\vspace{-0.5em}
The code that supports the findings of this study is available from the corresponding author upon request.

\renewcommand{\refname}{\large \sffamily References}
\bibliographystyle{unsrt}  
\bibliography{references}  







\end{document}